# Experimental observation of the steady – oscillatory transition in a cubic lid-driven cavity

A. Liberzon, Yu. Feldman and A. Yu. Gelfgat

School of Mechanical Engineering, Faculty of Engineering, Tel-Aviv University, Ramat Aviv, 69978, Tel-Aviv, Israel

**Abstract**

Particle image velocimetry is applied to the lid-driven flow in a cube to validate the numerical prediction of steady – oscillatory transition at lower than ever observed Reynolds number. Experimental results agree with the numerical simulation demonstrating large amplitude oscillatory motion overlaying the base quasi-two-dimensional flow in the mid-plane. A good agreement in the values of critical Reynolds number and frequency of the appearing oscillations, as well as similar spatial distributions of the oscillations amplitude are obtained.

## I. INTRODUCTION

Accurate prediction of the flow conditions in the driven cavity is of outmost importance for a number of technological applications, such as coating and polishing processes in microelectronics, passive and active flow control using blowing/suction cavities and riblets[1].Moreover, citing Shankar and Deshpande[1]"…the overwhelming importance of these flows is to the basic study of fluid dynamics". The driven cavity flow has well defined boundary conditions and it is apparently straightforward to use this configuration for benchmarking of numerical and experimental studies of fluid flows. Moreover, it can be shown that upon geometrical similarity (the width to height and width to span ratios) a single dimensionless number describes the flow state, namely the Reynolds number. Typically it is based on the cavity length, *L* and the driving lid velocity, *U*. The fixed flow domain makes this flow attractive also for experimental purposes, in particular, for studying flow transitions at large Reynolds numbers.Thoughapparently simple, the lid-driven cavity flows exhibit a vast variety of flow patterns, from 2D to 3D, secondary, corner and streamwise eddies, chaotic trajectories and more[1].

Despite the extensive research effort and very accurate numerical description of the two-dimension lid-driven cavity flow, the prediction of properties ofcorresponding three-dimensional flowsfor a given cavity geometry and at a given Reynolds number is still elusive. For example, the critical Reynolds number for a primary flow bifurcation was reported to be below 6000 (e.g. Ref. 1 and references therein). Bogatyrev&Gorin[2]and later Koseff& Street[3]experimentally observed 3-D unsteady flows at much lower Reynolds numbers,being close to 3000 in a cubic cavity. The observation was then verified by a more recent numerical study of Iwatsu et al.[4] who predicted an instability onset at the range of 2000 <$Re$< 3000. Another example is Figure 20a in Shankar and Deshpande[1]showing oscillations of velocity close to the wall of a cubic cavity, measured at $Re$ = 3200 by Prasad and Koseff.[5]The authors, however, did not pursue this research to identify the lowest Reynolds number at which the large amplitude fluctuations are observable. Very recently we (Feldman and Gelfgat[6]) reported rather accurate time-dependent computations of a 3D cubic lid-driven cavity flow, in which the steady – oscillatory transition was found to take place at even lower Reynolds number of approximately 1900. In order to validate these numerical results we conducted a series of experiments in a cubic cavity that was used for the Lagrangiantracking[7,8]. We report here on a good qualitative and quantitative agreement between the numerical and experimental results.

It is already a common knowledge that correct numerical prediction of instability onset requires correct resolution of both the base flow and the most unstable perturbation, which makes it a more challenging task than calculation of the flow only. Therefore, the agreement between the experiment and numerics observed for steady flows and for transition to unsteadiness yields a quite thorough validation of numerical results and predictions. Consequently, further analysis of, e.g. three-dimensional flow and disturbance patterns can be done on the basis of computational modeling, which is considerably simpler and more accurate than performing of corresponding measurements.

The experimental setup is briefly described in the following Section II, for the sake of completeness. We proceed to the main results in Section III, comparing the flow before and after the steady oscillatory transition. We present flow snapshots, describe in details the unstable modes and give some conclusive remarks in Section IV.

**II.EXPERIMENTAL SET UP**

Experiments are performed in twocubic cavities with a side length $L$= 40 and 80 mm,whoseupper boundary moves with a constant velocity $U$ in the $x$ direction as shown in Figure 1-a. All other cavity boundaries are stationary.The cavity is filled with a tap water and its moving lid comprises a circularly closed plastic belt driven by a DC motor.

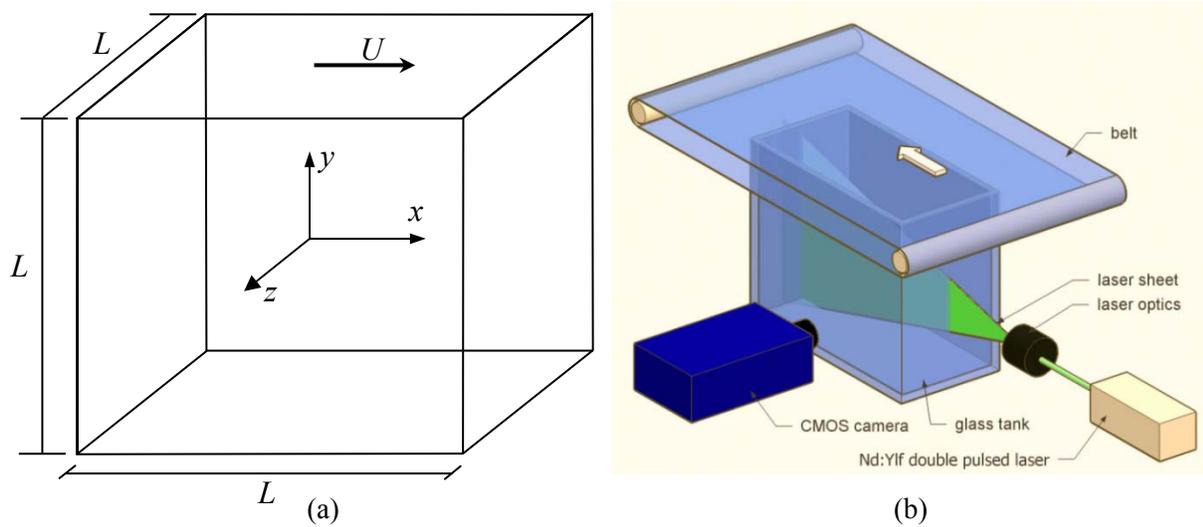

FIG. 1. Lid driven cavity: (a) physical model and coordinate system ;(b) sketch of the experimental set up.

The particle image velocimetry (PIV) technique was used for the flow measurements. The experimental setup is shown schematically in Figure 1-b.Amore detailed description including hardware and software components for data acquisition and processing, as well as estimated accuracy of the experimental measurements can be foundin Refs. [7,8]. The flow velocities were measured in the mid-plane, being alsothe symmetry plane, using particle image velocimetry (PIV) system by TSI Inc. (including the 120 mJNewWave Solo Nd:YAG laser, 4096 x 2048 pixels CCD camera, Nikkor 60mm lens). About 2000 PIV snapshots weretaken at the rate of 2 Hz for $L$=80mm and 15 Hz for $L$=40mm (applying Mikrotron MC1324 10 bit, 1280 x 1024 pixels CMOS camera). The time-dependent data was analyzed using the standard FFT-based cross-correlation algorithms using commercial (Insight 3G ver. 9.10, TSI Inc.) and open-source (OpenPIV ,http://www.openpiv.net) software forthe verification purposes.

Experimental uncertainty of a single PIV velocity realization is estimated to be less than 5% relative to the full-scale. This estimation is based on the standard error analysis. Among other routine checks we verify that peak locking is not present in our PIV measurements.[9] Thus, the uncertainty of the ensemble average in the central part of flow field is below 0.1%.Apparently,additional experimental uncertainties related to the measurement of the belt velocity, determination of the fluid viscosity,measurement of the cavity dimensions and variations of the environment temperaturealways exist. However, all of them are at least order of magnitude less than the PIV error in the near-wall regions which bounds the estimation of overall measurement uncertaintyto be of the order of 5% for the averaged values. The main reason for this error is the reflection from the glass walls and from the belt.

### III. EXPERIMENTAL RESULTS AND DISCUSSION

All experimental results obtained in the present study were normalized using the scales*L*, *U*, *t*=*L/U* for length, velocity and time,respectively. Thus,the only dimensionless parameter determining the flow in a cubic cavity is the Reynolds number defined as *Re*=*UL*/$\tilde{\nu}$, where $\nu$ is a kinematic viscosity.Present experiments were performed for two different cavity lengths and two different working liquids.In each case the lid velocity was varied to obtain the stable flow below and above the critical point, predicted in our numerical study (Feldman and Gelfgat[6]).By variation of experimental liquids and the cavity size we change both the viscous time scale ($L^2/\nu$) and kinematic time scale ($U/L$), preserving the same value of the *Re* number and verifying that at the observed flow oscillations are not induced by the experimental setup, e.g., by the return-period of the belt. In each case no qualitative effect on the main dimensionless frequency of oscillations of slightly supercritical flows and on the oscillations amplitude distribution in the cavity midplane had been observed (see below).

*Validation of the experimental results*

Figure 2 illustrates a cross-verification of the experimental and numerical results in the sub-critical steady state regime, showing velocity distributions along two centerlines (*x*,0,0) and (0,*y*,0) at the cavity mid-plane cross section for *Re*= 1480. A very good agreement is observed between the experimental and numerical results for the entire range of $v_x$ and $v_y$ velocity components with the exception of two narrow regions -0.5 ≤ *y* ≤ -0.4 and 0.4 ≤ *y* ≤ 0.5 where the PIV measurements are not as accurate as in the central region. The overall agreement of the experimental data and computational analysis, taking into account the imperfections of the experimental setup, is 4.3% within the 5% uncertainty.

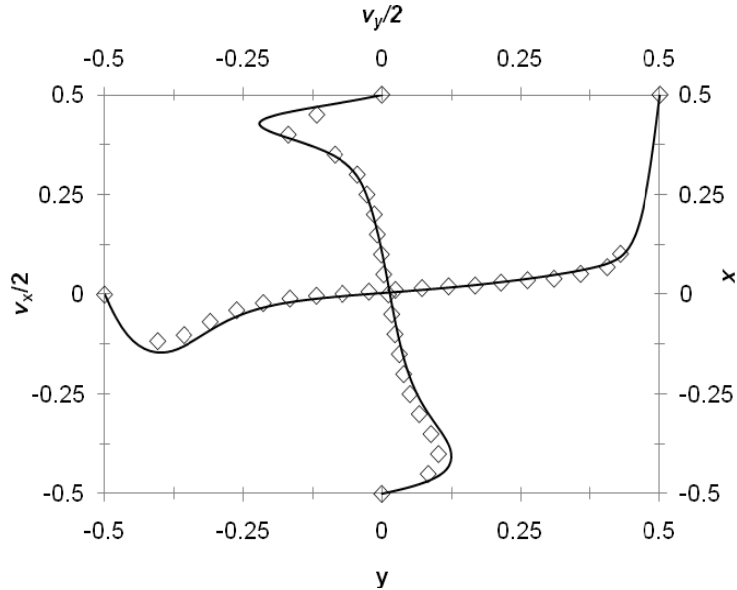

FIG.2. Comparison of the numerically (solid line) and experimentally (◊) obtained centerline velocities at the cavity mid-plane, *Re* = 1480 (*L*=80mm).

*Steady and unsteady flow analysis*

According to the numerical results of Feldman & Gelfgat,[6] who investigated a set of sub- and slightly supercritical flows in a cubic lid-driven cavity, the steady – unsteady transition occurs via subcritical Hopf bifurcation with $Re_{cr} \approx 1914$. It was also observed that at *Re* = 1970 the flow exhibits oscillations characterized by a dimensionless angularfrequency ω=0.575 and its subsequent multiple harmonics. This fact motivated the next step of the present study at which a set of experiments have been performed for increasing Reynolds numbers: $Re$ =1480, 1700, 1970, 2100 in both cavities. Two first series of experiments were carried out in the cavity with the side length of 80mm using water and glycerin solution as working liquids. Then the experiment was repeated in a smaller cavity with the length size of 40mm filled with water. For estimation of Reynolds number the water kinematic viscosity was taken $10^{-6}$ m$^2$/s, and of glycerin solution $1.68 \times 10^{-6}$ m$^2$/s. It should be emphasized that because of the assumed sub-criticality of the Hopf bifurcation[6] the experiments were performed in order of increasing $Re$ numbers. When going between two adjacent $Re$ numbers, the data acquisition for each $Re$ started only after a sufficiently long time period (at least 500 turn-over times) necessary for the flow to reach asymptotic steady or oscillatory state. Figure 3 presents characteristic oscillations of $v_x$ measured for three successive $Re$ numbers after all the transient changes already took place. Figure 4 presents the corresponding Fourier spectra. The results are shown for the points with dimensionless coordinates (-0.325, -0.378, 0) for water and (-0.342, -0.395, 0) for glycerin solution, where observed oscillations amplitude is close to the maximal one for all values of the Reynolds number. To make the comparison easier the oscillations, frequencies and amplitudes are shown after being rendered into dimensionless values.

At $Re$=1480 we observe only system noise as it clearly seen from the signals (Figure 3) and the spectra (Figure 4-a). It is noteworthy that in experiment with water, whose viscosity is smaller, the noise has larger amplitude. With the increase of Reynolds number up to $Re$=1700 we already observe oscillations characterized by a weak dominant single-frequency (Figure 4-b). The frequency is close to those observed for well-developed oscillatory flows at larger $Re$, so that we undoubtedly observe oscillations of the dominant eigenmode of the flow. On the other hand, according to our numerical predictions[6] the flow at this Reynolds number is expected to exhibit slowly decaying oscillations of the dominant instability mode. Clearly, the experimental setup is far from being ideal reproduction of the mathematical model. Possibly, the finite-amplitude experimental noise excites the dominant eigenmode during the experiment, however its stable nature does not allow it to grow. This assumption is supported also by observation of larger dimensionless amplitudes in water where the viscosity is smaller.

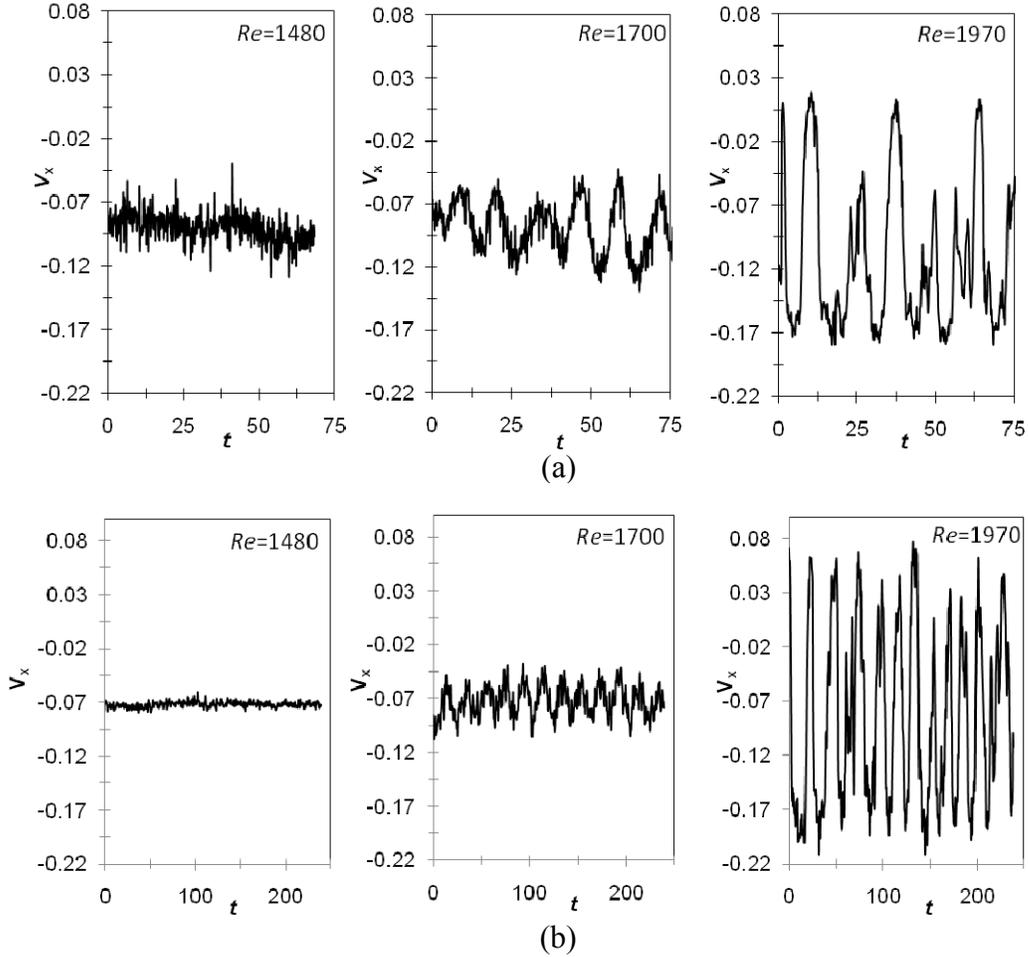

FIG.3 Dimensionless time evolution of the $v_x$ velocity component for different Re numbers at the cavity midplane: (a) control point (-0.325,-0.378, 0) for water; (b) control point (-0.342, -0.395, 0) for glycerin solution. Cavity with the side length of 80mm.

Further increase of the Reynolds number up to $Re$=1970 leads to the flow oscillations with a significantly larger amplitude. Note that dimensionless amplitudes of the main harmonics are close for water and glycerin solution experiments, thus indicating that we observe an asymptotic oscillatory state rather than noise-induced oscillations. The differences in spectra and the frequencies of main harmonics need a special discussion (see below). Here we emphasize that when the Reynolds number is increased to $Re$=2160 the dimensionless amplitudes of the main harmonics reduce, while many other low-amplitude harmonics appear in the spectra. This shows that the value *Re*=2160 is significantly far from a single-frequency asymptotic state expected to be observed as a result of the primary instability of the steady flow.

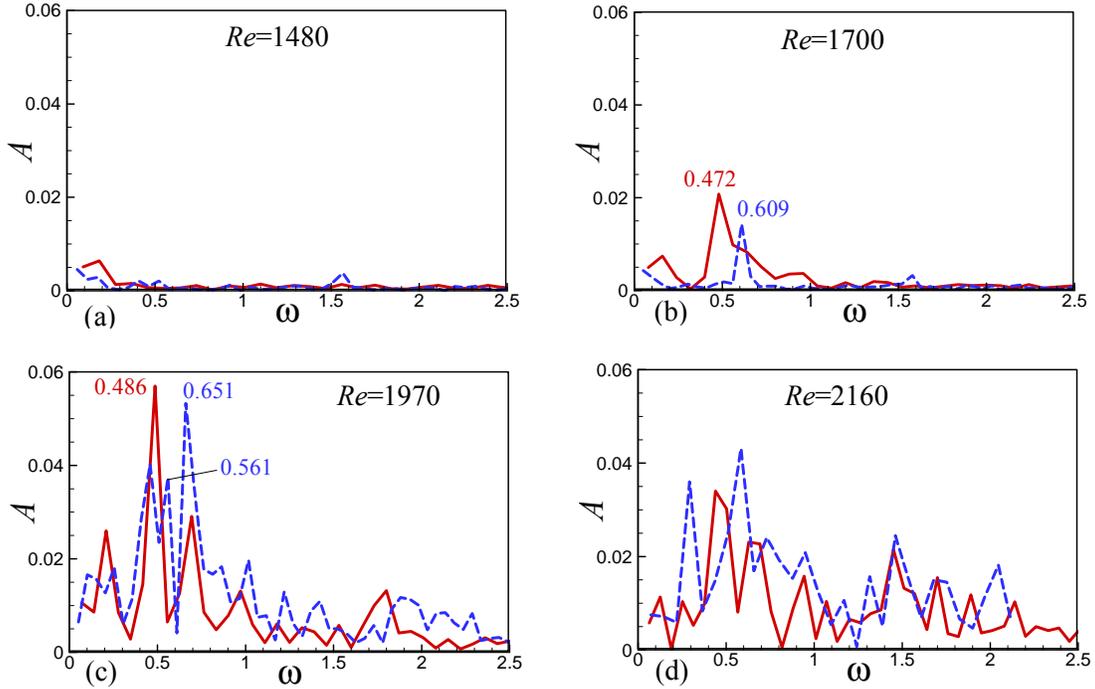

FIG.4. Fourier transform of $v_x$ velocity component measured in the cavity mid-plane (z = 0). Solid red and dash blue lines correspond to experiments with water and glycerin solution, respectively. Cavity with the side length 80 mm.

The main dimensionless frequencies observed for experiments with water and glycerin solution are, respectively, 0.486 and 0.651. Note, that they are smaller or larger the numerically predicted value 0.575 by approximately same increment. The discrepancies in measured and calculated frequencies, as well as presence of other frequencies in the spectra (Figure 4-c) that were not observed numerically can be explained by several experimental imperfections. First, among some high-frequency vibrations caused by the motor, the system contains an internal period of excitation connected with the period of the belt motion. As stated above, to ensure that the instability observed is not related to this period we performed an additional experiment using the same belt and a smaller cavity with the side length 40mm, thus making the difference between the belt and flow periods significantly larger. The main dimensionless frequency observed in a smaller cavity at $Re$ = 1970 was 0.61, which is well compared with the above results. The latter frequency is closer to the theoretically predicted value 0.575, which shows that flow in a smaller cavity is less affected by the system oscillations. It is emphasized, however, that unavoidable presence of oscillations makes the flow parametrically excited, so that the whole system becomes qualitatively different from the classical mathematical model. This, in particular, does not allow us to localize critical Reynolds number more precisely.

The second unavoidable experimental imperfection is the contact between the belt and the working liquid. Really, in the mathematical formulation[6] the boundary conditions for *x*-velocity in the upper corners are discontinuous. Clearly, this cannot be reproduced in the experimental setup, where boundary conditions in these corners are not very well controlled and may change when the cavity is open and closed again. Obviously, such a boundary conditions imperfection can affect the flow spectrum and excite some additional frequencies close to one unstable in the purely ideal case. We believe that these two reasons are mainly responsible for the slight disagreement with the numerical predictions, as well as for the scattering of the experimental results themselves.

For further comparison with the numerical results and to ensure that observed oscillations result from the same instability mechanism as the numerical predictions we filter out the main harmonics of oscillations using a standard band-pass filter. Figure 5 compares spatial distributions of $v_x$ and $v_y$ amplitudes at the cavity mid-plane computed and measured at $Re$=1970. It is clearly seen that the spatial distribution of the main frequency amplitudes predicted by computational modeling is fully represented by recent measurements both in water and glycerin solution. This striking qualitative agreement between the numerical predictions and the experimental observations made for both amplitudes makes us confident of the fact that the same instability has been computed and measured. As it has been already stated in Ref. [6] the maximum values of $v_x$ and $v_y$ oscillation amplitudes are located on the border between the primary eddy with the secondary downstream and upstream eddies located respectively in the lower right and left corners of the mid plane cross section (see also Figure 6).

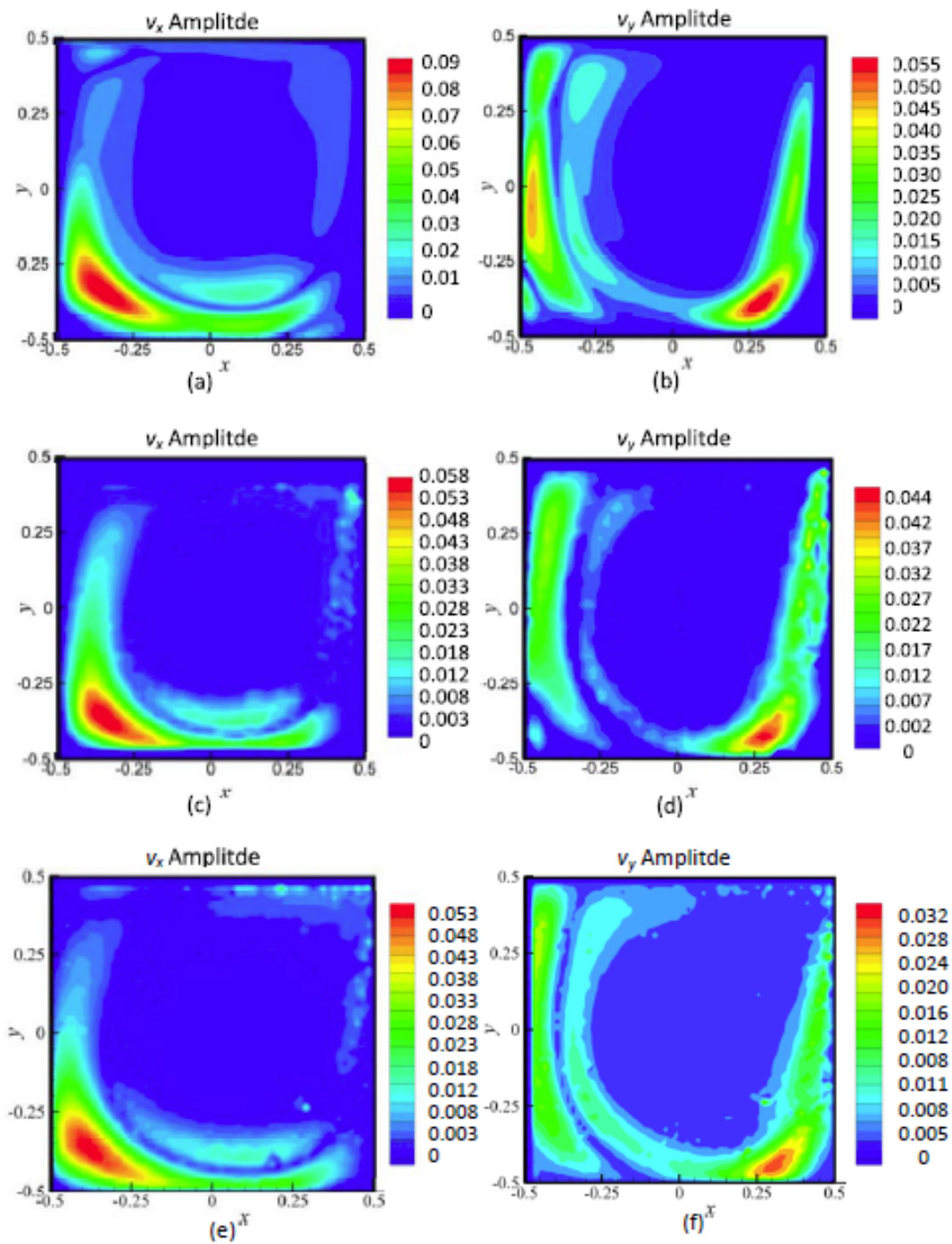

FIG.5. Spatial distributions of maximal $v_x$ and $v_y$ oscillation amplitudes at the cavity mid-plane, $Re = 1970$: (a)-(b) numerical results; (c)-(d) experimental results for water; (e)-(f) experimental results for glycerin solution.

The quantitative comparison revealsthat the numerical values of $v_x$and $v_y$amplitudes are larger than the experimental ones.The maximum values of the measured$v_x$ and $v_y$ amplitudes comprise about 64% and 80%, respectively, from the corresponding computed values. This fact may be explained by existence of the energy dispersion intrinsic in the experimental set up(see above) and also can be connected tothe measurementinaccuracies near the walls. The latter explains alsothe deviationsbetween the numerical and the experimental spatial locations at which the maximal amplitude values of both velocity components are observed (see Table I).

Table I. Comparison between locations of maximal values of velocity amplitudes obtained numerically experimentally.

|  | Maximal value of $v_x$ Amplitude | Maximal value of $v_y$ Amplitude |
|---|---|---|
| Numerical results | (−0.338,−0.343,0) | (0.289,−0.383,0) |
| Experimental results, water | (−0.325,−0.378,0) | (0.275,−0.427,0) |
| Experimental results, glycerin solution | (−0.342, −0.395, 0) | (0.307,−0.447,0) |

We use the filtered flow field, containing only the most unstable mode and its subsequent doubled harmonics,for avisualization of the 3D cavity flow. Figure 6–a presents a snapshot of the typical flow pattern at the cavity mid-plane at$Re$ =1970 characterized by secondary upstream and secondary downstream eddies located in the left and right corners, respectively, and by a primary eddy located in the central part of the cavity. Figures 6-b-e showfour velocity snapshotsat the left and right corners of the mid-plane where the maximum values of velocity oscillations are observed. It is assumed that the latter, as well as the entire instability mechanism, is a result of interaction between the primary and the secondaryeddies taking place at theborderseparating them.The oscillating pattern of a slightly supercritical flow has been already described numerically in Ref. 6. The snapshots in Figure 6 are equally spaced in time over a single period. For the visualization purposes all velocity vectors are plotted with the uniform length independently of their numerical values. Because of the low precision of the flow properties measured close to the cavity boundaries the velocity values in these regions were estimated by a linear interpolation between the corresponding boundary values, known from the non-slip boundary conditions, and the nearby measured interior velocity values. The observed oscillations of the corner vortices qualitatively agree with the numerical predictions of Ref. 6.

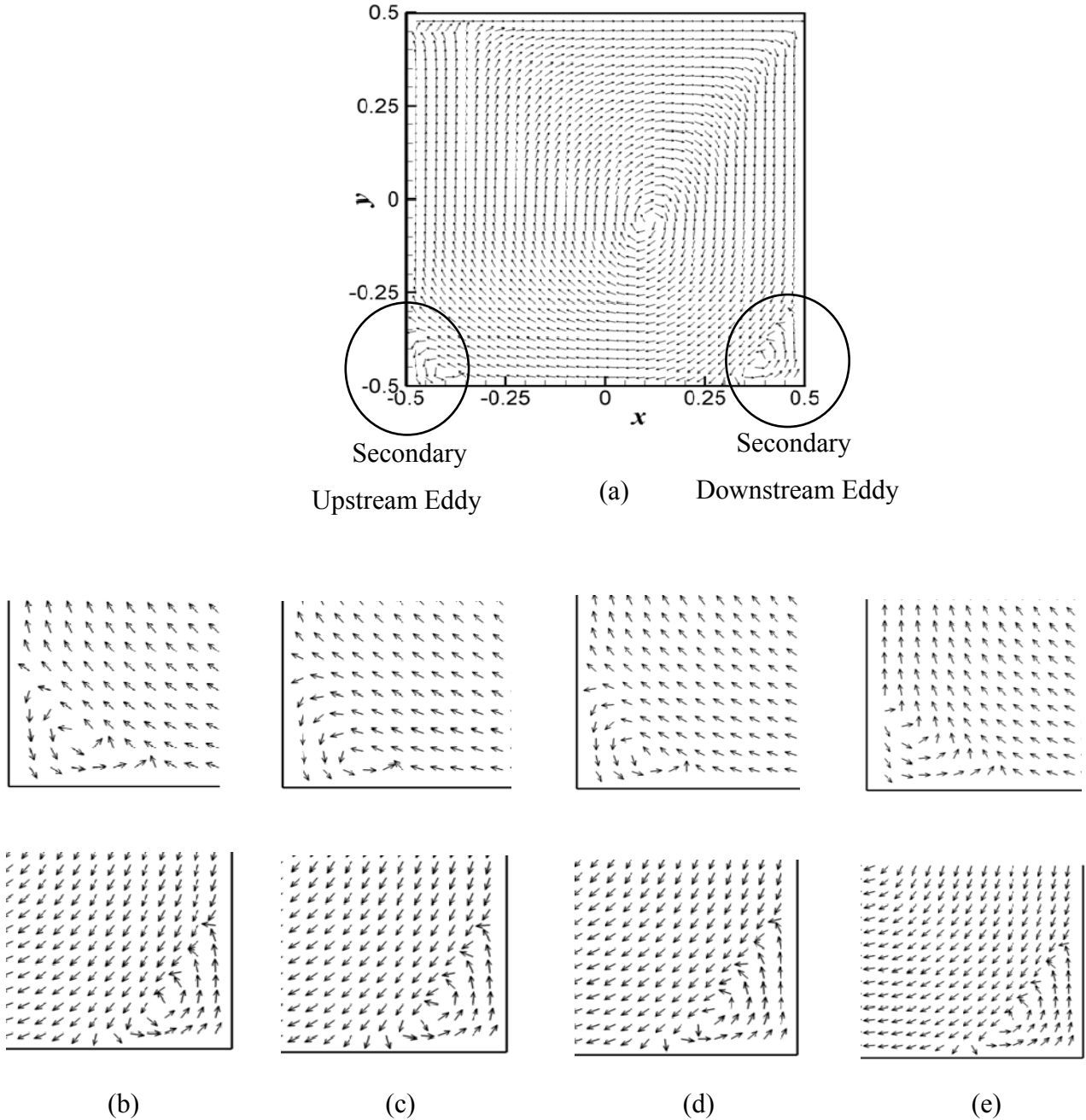

FIG. 6. Flow velocity vectors at the cavity mid-plane $(x, y, 0)$ over a single period, $Re = 1970$. Experiment with water as working liquid. (a) General view; (b)-(e) velocity snapshots of the secondary upstream and downstream eddies: (b) $t = 0$; (c) $t = 3.23$; (d) $t = 6.46$; (e) $t = 9.7$. (Enhanced online)

## IV. CONCLUSSIONS

An experimental study of a cubic lid driven cavity flow for a set of *Re* numbers corresponding to the steady and unsteady flow regimes has been performed. The study showed that the steady state is stable for*Re*<1700 at least. The obtained experimental results have been then successfully compared with the corresponding numerical steady state solutions. It was found thata steady – unsteady transition occurs in the range1700 <*Re*< 1970.Beyond *Re*=1970 the flow becomes oscillatory. Both, location of the threshold and the measured oscillations frequencies are in a good agreement with the numerical results of Ref. 6where instability had been predicted at $Re_{cr} \approx 1914$ withω≈0.575. The experimental patterns of the spatial distribution of the velocity amplitudes are in good qualitative agreement with that of the numerical solution that allows us to arguethat the experimental and numerical observations resultfrom the same instability mechanism. An accurate quantitative measurement of the critical Reynolds number remainsopen and challenging issue. Nevertheless, we believe that the accuracy of the present results is sufficient for the validation of our recent numerical prediction.[6]Namely,the experimentally observed instability sets in at Reynolds numbers that aresignificantly lower than those predicted in former experiments and rather close to the recentnumerically predicted value.

**Acknowledgement**

This study was supported by German-Israeli foundation, grant No. I-954 -34.10/2007, and by the Israel Science Foundation, grant No. 782/08.